\documentclass[a4paper, 11pt]{article}
\pdfoutput=1
\usepackage{jheppub}

\usepackage[caption=false]{subfig}
\usepackage{amsfonts}
\usepackage{textcomp}
\usepackage{gensymb}
\usepackage{booktabs}

\usepackage[utf8]{inputenc}

\usepackage{listings}
\lstdefinestyle{mystyle}{
    basicstyle=\ttfamily,
    breakatwhitespace=false,         
    breaklines=true,                 
    captionpos=b,                    
    keepspaces=true,                 
    showspaces=false,                
    showstringspaces=false,
    showtabs=false,                  
    tabsize=4
}
\AtBeginDocument{
\heavyrulewidth=.08em
\lightrulewidth=.05em
\cmidrulewidth=.03em
\belowrulesep=.65ex
\belowbottomsep=0pt
\aboverulesep=.4ex
\abovetopsep=0pt
\cmidrulesep=\doublerulesep
\cmidrulekern=.5em
\defaultaddspace=.5em
}

\newcommand{\like}{\mathcal{L}}
\newcommand{\vectheta}{\vec{\theta}}
\newcommand{\vecw}{\vec{w}}
\newcommand{\prob}{\mathcal{P}}
\newcommand{\gprob}{\mathcal{G}}
\newcommand{\meanl}{\mathcal{L}_{\textmd{Mean}}}
\newcommand{\mcl}{\like_\textmd{Eff}}
\newcommand{\gl}{\like_\textmd{G}}
\newcommand{\adhoc}{\mathcal{L}_{\textmd{AdHoc}}}
\newcommand{\lpoisson}{l_{\textmd{Poisson}}}
\newcommand{\lmc}{l_\textmd{Eff}}
\newcommand{\lbarlow}{\like_{\textmd{BB}}}
\newcommand{\hatmu}{\hat{\mu}}
\newcommand{\hatpoisson}{\hatmu_\textmd{Poisson}}
\newcommand{\hatmc}{\hatmu_\textmd{Eff}}

\newcommand{\agpar}{\alpha}
\newcommand{\bgpar}{\beta}
\newcommand{\emcee}{\texttt{emcee}}
\newcommand{\meff}{m_\mathrm{Eff}}
\newcommand{\weff}{w_\mathrm{Eff}}

\usepackage{tikz}
\usepackage{environ}

\usetikzlibrary{calc,trees,positioning,arrows,chains,shapes.geometric,%
    decorations.pathreplacing,decorations.pathmorphing,shapes,%
    matrix,shapes.symbols,backgrounds,fit} % required in the preamble
\usepackage{varwidth}

\makeatletter
\newsavebox{\measure@tikzpicture}
\NewEnviron{scaletikzpicturetowidth}[1]{%
  \def\tikz@width{#1}%
  \begin{lrbox}{\measure@tikzpicture}%
  \BODY
  \end{lrbox}%
  \pgfmathparse{#1/\wd\measure@tikzpicture}%
  \BODY
}
\makeatother

\tikzset{
	>=stealth',
	box/.style={
		rectangle,
		rounded corners,
		dashed,
		draw=black, very thick,
		minimum height=2em,
		text centered,
		execute at begin node={\begin{varwidth}{28em}},
		execute at end node={\end{varwidth}}},
	solidbox/.style={
		rectangle,
		rounded corners,
		draw=black, very thick,
		minimum height=2em,
		text centered,
		execute at begin node={\begin{varwidth}{28em}},
			execute at end node={\end{varwidth}}},
    bigsolidbox/.style={
		rectangle,
		rounded corners,
		draw=black, very thick,
		minimum height=6cm,
		text centered,
		execute at begin node={\begin{varwidth}{28em}},
			execute at end node={\end{varwidth}}},
	fw_arrow/.style={
		->,
		thick,
		shorten <=2pt,
		shorten >=2pt,},
	bw_arrow/.style={
		<-,
		thick,
		shorten <=2pt,
		shorten >=2pt,}
}

\definecolor{mc_gen_color}{RGB}{250,138,31}

\definecolor{det_sim_color}{RGB}{227,66,55}

\definecolor{llh_color}{RGB}{128,0,128}

\tikzstyle{bigboxGeneration} = [draw=mc_gen_color!50, thick, fill=mc_gen_color!20, rounded corners, rectangle]

\tikzstyle{bigboxDetector} = [draw=det_sim_color!50, thick, fill=det_sim_color!20, rounded corners, rectangle]

\tikzstyle{bigboxAnalysis} = [draw=llh_color!50, thick, fill=llh_color!20, rounded corners, rectangle]

\title{A binned likelihood for stochastic models}

\author[a,1]{C.A.~Arg\"uelles, \note{ORCID: \href{https://orcid.org/0000-0003-4186-4182}{0000-0003-4186-4182}}}
\author[b,2]{A.~Schneider, \note{ORCID: \href{https://orcid.org/0000-0002-0895-3477}{0000-0002-0895-3477}}}
\author[b,3]{T.~Yuan, \note{ORCID: \href{http://orcid.org/0000-0002-7041-5872}{0000-0002-7041-5872}}}
\affiliation[a]{Dept.~of Physics, Massachusetts Institute of Technology, Cambridge, MA 02139, USA}
\affiliation[b]{Dept.~of Physics and Wisconsin IceCube Particle Astrophysics Center, University of Wisconsin, Madison, WI 53706, USA}
\emailAdd{caad@mit.edu}
\emailAdd{aschneider@icecube.wisc.edu}
\emailAdd{tyuan@icecube.wisc.edu}

\abstract{Metrics of model goodness-of-fit, model comparison, and model parameter estimation are the main categories of statistical problems in science. Bayesian and frequentist methods that address these questions often rely on a likelihood function, which is the key ingredient in order to assess the plausibility of model parameters given observed data. In some complex systems or experimental setups, predicting the outcome of a model cannot be done analytically, and Monte Carlo techniques are used. In this paper, we present a new analytic likelihood that takes into account Monte Carlo uncertainties, appropriate for use in the large and small sample size limits. Our formulation performs better than semi-analytic methods, prevents strong claims on biased statements, and provides improved coverage properties compared to available methods.}

\begin{document}
\maketitle

\section{Introduction\label{sec:intro}}

The use of Monte Carlo (MC) techniques to calculate nontrivial theoretical quantities and expectations in complex experimental settings is common practice in particle physics.
A MC event is a single representation of what can be detected in data and is typically generated from a single realization of the underlying physics parameters, $\vectheta_g$. These events are often binned in some observable space and compared with the data. Since the generation process is stochastic, a particular $\vectheta_g$ used for generating the MC can lead to different outputs. This stochasticity introduces an uncertainty in the MC distributions. Furthermore, as production of large MC is often time-consuming, reweighting is used to move from one hypothesis to another. In reweighting, each MC event is assigned a new weight, $w(\vectheta)$ that accounts for the difference between the generation parameters $\vectheta_g$ and the hypothesis parameters $\vectheta$~\cite{Gainer:2014bta}. It follows that MC uncertainties will be hypothesis dependent; thus, to do hypothesis testing, it is important to account for them. This is especially important for small-signal searches, performed in the small sample limit, where a modified-$\chi^2$ may not be suitable~\cite{Lyons:1986em}. A Poisson likelihood is a more appropriate statistical description of event counts~\cite{poisson1837recherches}, but in that case a proper treatment of MC statistical uncertainties is less straightforward. Solutions to this problem have been discussed in the literature in the context of frequentist statistics by adding nuisance parameters~\cite{Barlow:1993dm,Cranmer:2012sba,Chirkin:2013lya}, as well as detailed probabilistic treatment of MC weights~\cite{Glusenkamp:2017rlp}. However, \cite{Barlow:1993dm, Chirkin:2013lya, Glusenkamp:2017rlp} add additional time complexity, and \cite{Cranmer:2012sba} does not provide a full exposition on how to incorporate weighted MC. We present a new treatment that is valid in the large and small limit of the data sample size, suited for frequentist and Bayesian analyses, based on the Poisson likelihood. Our likelihood accounts for statistical uncertainties due to MC, allows for arbitrary event-by-event reweighting, and is computationally efficient. A test statistic based on the proposed likelihood is found to follow a distribution closer to the asymptotic form expected from Wilks' theorem. An implementation of the likelihood described in this work can be found in~\cite{MCLLH}.

This paper is organized as follows. In Sec.~\ref{sec:mc_intro} we briefly review two common treatments available in the literature to account for MC statistical uncertainty. In Sec.~\ref{sec:generalization_poisson} we define and discuss our new likelihood. In Sec.~\ref{sec:example} we study the performance of the likelihood through an example and compare it to other likelihoods in the literature. In Sec.~\ref{sec:conclusion} we provide our conclusions. A summary of the likelihoods discussed in the paper, including our main result, is given in Appendix~\ref{sec:appendixA}.

\section{The Poisson likelihood and previous work\label{sec:mc_intro}}

In order to compare MC with data, events are often binned into distributions across a set of observables. For simplicity we focus on a single bin. In the absence of cross-bin-correlated systematic uncertainties the generalization to multiple bins is simply a product over the likelihood in all bins. This is assumed for the remainder of the paper. It is well known that the count of independent, rare natural processes can be described by the Poisson likelihood, given by
\begin{equation}
\label{eq:poisson}
\like(\vectheta|k) = \mathrm{Poisson}(k;\lambda(\vectheta)) = \frac{\lambda(\vectheta)^{k}e^{-\lambda(\vectheta)}}{k!},
\end{equation}
where $\lambda(\vectheta)$ is the expected bin count for a hypothesis and $k$ is the number of observed data events. Equation \eqref{eq:poisson} requires exact knowledge of the expected bin count, $\lambda(\vectheta)$. In the case of complex experiments it is often not possible to obtain $\lambda(\vectheta)$ exactly and MC techniques are used to estimate the expected distributions. For weighted MC, often a direct substitution of $\lambda(\vectheta)$ by $\sum_i{w_i(\vectheta)}$ is used, where $w_i$ are the weights of each of the MC events in the bin. Then Eq.~\eqref{eq:poisson} can be approximated as
\begin{equation} \label{eq:mcpoisson}
\adhoc(\vectheta|k) = \frac{\left(\sum_{i}{w_i(\vectheta)}\right)^{k}e^{-\left(\sum_{i}{w_i\left(\vectheta\right)}\right)}}{k!}.
\end{equation}
This ad hoc treatment assumes that the MC estimate of the expected bin counts exactly matches the true expectation rate of the model, neglecting the stohastic nature of MC. In the case of large MC, Eq.~\eqref{eq:mcpoisson} converges to Eq.~\eqref{eq:poisson} for the hypothesis given by $\vectheta$.

\subsection{The Barlow-Beeston likelihood}
To treat MC statistical uncertainties in the small sample limit, a modification of the Poisson likelihood was introduced in~\cite{Barlow:1993dm}, which is briefly covered below. First, note that the expectation in a single bin is given by contributions from different physical processes, which we index by $j$. Then, the number of expected events can be written as
\begin{equation}
\lambda(\vectheta) = \sum_{j=1}^s \bar n_{j}(\vectheta),
\label{eq:barlow_true_nonweighted}
\end{equation}
where $\bar n_{j}$ is the expected number of MC events from process $j$ that fall in the bin and $s$ is the total number of relevant processes. Substituting Eq.~\eqref{eq:barlow_true_nonweighted} into Eq.~\eqref{eq:poisson} gives the Poisson likelihood for observing $k$ data events. For stochastic models, $\bar n_j$ is unknown. Instead, the MC outcome can be modeled as having drawn $n_j$ events from a random process that simulates the physical process. When MC generation is expensive, we can approximate $n_j$ as being drawn from a Poisson process with mean $\bar{n}_j$~\footnote{The MC generation is a binomial process where we generate a fixed number of events for each process, $N_j$, and accept them into the bin of interest with probability $\beta_{j}(\vectheta)$, such that $\bar n_{j}(\vectheta)=\beta_{j}(\vectheta) N_j$. In the limit of both of rare processes ($\beta_{j} \ll 1$) and large number of generated events ($N_j \gg 1$), the total number of observed events can be approximated as Poisson distributed with mean $\lambda(\vectheta) = \sum_j\beta_{j}(\vectheta) N_j = \sum_j\bar n_{j}(\vectheta)$.}. Profiling on the true number of MC events per process in the bin results in the Barlow-Beeston (BB) likelihood, given by~\cite{Barlow:1993dm}
\begin{equation}
\lbarlow(\vectheta|k) = \underset{\{\bar n_j\}}{\rm max}
\frac{\lambda(\vectheta)^{k}e^{-\lambda(\vectheta)}}{k!} \prod_{j=1}^s \frac{\bar n_j^{n_j}e^{-\bar n_j}}{n_j!},
\label{eq:barlow_likelihood_nonweighted}
\end{equation}
where $\lambda(\vectheta)$ is given by Eq.~\eqref{eq:barlow_true_nonweighted}, $n_{j}$ and $\bar{n}_{j}$ are the estimated and true MC counts in the bin respectively, and $\{\bar{n}_j\}_{j=1}^{s}$ denotes the $s$ nuisance parameters we have profiled over.

In the above formalism we have produced the MC at the natural rate, but this is not the case for weighted MC. The prescription is given by replacing Eq.~\eqref{eq:barlow_true_nonweighted} with
\begin{equation}
\lambda(\vectheta) = \sum_{j=1}^s \eta_{j}(\vectheta) \bar n_j,
\label{eq:barlow_true_weighted}
\end{equation}
where $\eta_{j}(\vectheta)$ is a scale factor for process $j$ that accounts for the differences in the MC generation and the target hypothesis of interest. In this case, the likelihood definition is still given by Eq.~\eqref{eq:barlow_likelihood_nonweighted}; an explicit formula for $s=1$ is given in the appendix. However, for arbitrary weight distributions per physical process $\lbarlow$ may not be appropriate as it neglects the variance from a sum of weights~\cite{Barlow:1993dm}. It remains valid only in the case where the distribution of weights for each process is narrow.

\subsection{Uncertainties in the large-sample limit}
In the large-sample regime, the Gaussian distribution is an appropriate description of the observed data. In this limit, the use of Pearson's $\chi^2$ as a test-statistic~\cite{Pearson:1900} is common practice. For a single analysis bin, Pearson's $\chi^2$ is defined as
\begin{equation}
\chi^2(\vectheta) = \frac{(k - \lambda(\vectheta))^2}{\lambda(\vectheta)},
\label{eq:chi2_pearson}
\end{equation}
where we continue to use the approximation $\lambda(\vectheta) = \sum_i{w_i(\vectheta)}$ and $w_i$ are the weights of each of the MC events. The form of Pearson's $\chi^2$ arises from the fact that the Gaussian distribution of $k$ is the large-sample limit of a Poisson distribution for which the expected statistical variance of the observation is given by $\lambda(\vectheta)$. Systematic uncertainties, under the assumption that they follow a Gaussian distribution and are independent between bins, can be included as
\begin{equation}
\chi^2(\vectheta) = \frac{(k - \lambda(\vectheta))^2}{\lambda(\vectheta) + \sigma^2_{\rm syst.}}.
\label{eq:chi2_systematics}
\end{equation}
However, this method of incorporating systematic uncertainties tends to overestimate them in shape-only analyses; see~\cite{Cogswell:2018auu} for a recent discussion in the context of reactor neutrino anomalies. Similarly, one can include uncertainties to account for statistical fluctuations of the MC in the test-statistic. In doing so, the Gaussian behavior is implicit and the modified $\chi^2$ reads
\begin{align}
\chi^2_{\rm mod}(\vectheta) = \frac{(k - \lambda(\vectheta))^2}{\lambda(\vectheta) + \sigma^2_{\rm syst.} + \sigma^2_{\rm mc}},
\label{eq:modified_chi2}
\end{align}
where $\sigma^2_{\rm mc}$ is the MC statistical uncertainty in the bin given by
\begin{equation}\label{eq:sigma}
\sigma^2_{\rm mc}(\vectheta) \equiv \sum_{i=1}^m w_i(\vectheta)^2.
\end{equation}
Note that this test-statistic definition is not appropriate in the small-sample regime, as the data is no longer well described by a Gaussian distribution. If one uses a $\chi^2$ test-statistic in the small-sample regime, one ought to calculate the test-statistic distribution properly to achieve appropriate coverage~\cite{cowan1998statistical}.

\section{Generalization of the Poisson likelihood\label{sec:generalization_poisson}}

%Using MC we can compute the expected event count, $\lambda(\vectheta)$, by means of adding the MC event weights, as discussed in Sec.~\ref{sec:mc_intro}. Due to the finite nature of MC, we can never recover exactly the intrinsic parameter, $\lambda({\vectheta})$, from which the MC events are drawn, but can only construct estimators; the same is true for any observable of interest. As such, $\adhoc$ written in Eq.~\eqref{eq:mcpoisson} is an approximation that presumes the MC realization matches the expectation.

Ideally we would like to obtain the expected event count for any hypothesis, $\lambda(\vectheta)$, however we are considering problems where this relationship is not known and $\lambda$ is instead estimated by MC. The key difference here is that instead of using exact knowledge of $\lambda$ we want to perform Bayesian inference to obtain $\prob(\lambda|\vectheta)$ using the MC available. Assuming the weights are functions of $\vectheta$, we have 
\begin{equation} \label{eq:generalpoisson}
\like_{\rm General}(\vectheta|k) = \int_{0}^{\infty}~\frac{\lambda^{k}e^{-\lambda}}{k!}\prob\left(\lambda|\vecw(\vectheta)\right)~d\lambda,
\end{equation}
where the distribution of $\lambda$, $\prob\left(\lambda|\vecw(\vectheta)\right)$, is inferred from the MC. %Note that to obtain $\like_{\rm General}$ we have implicitly made the assumption that $\prob(w_i'|\vectheta)=\delta(w_i'-w_i(\vectheta))$ for each event $i$ in the bin, and integrated over all possible $\vec{w_i'}$.
The likelihood, $\adhoc$, in Eq.~\eqref{eq:mcpoisson} is recovered when $\prob(\lambda|\vecw(\vectheta)) = \delta\left(\lambda - \sum_{i}{w_i(\vectheta)}\right)$, but clearly this is an unrealistic assumption as it presumes perfect knowledge of the parameter $\lambda(\vectheta)$ from a finite number of realizations. Instead, it is more appropriate to construct $\prob(\lambda|\vecw(\vectheta))$ based on the MC realization. This is given by
\begin{equation} \label{eq:posterior}
\prob\left(\lambda|\vecw(\vectheta)\right) = \frac{\like(\lambda|\vecw(\vectheta))\prob(\lambda)}{\int_0^\infty \like(\lambda'|\vecw(\vectheta))\prob(\lambda')~d\lambda'},
\end{equation}
where $\prob(\lambda)$ is a prior on $\lambda$ that must be chosen appropriately and $\like(\lambda|\vecw(\vectheta))$ is the likelihood of $\lambda$ given $\vecw(\vectheta)$. This is similar to~\cite{Barlow:1993dm, Cranmer:2012sba}, but instead of fitting $\lambda$ as a nuisance parameter as in $\lbarlow$ in Eq.~\eqref{eq:barlow_likelihood_nonweighted}, we marginalize over it in Eq.~\eqref{eq:generalpoisson} as informed by the MC weights. When $\like_{\rm General}$ is used under a frequentist approach, the marginalization over $\lambda$ implies a hybrid Bayesian-frequentist construction, similar to the treatment of nuisance parameters described in~\cite{Cousins:1991qz} and employed in~\cite{Abe:2017vif, Abe:2018wpn}.

This section is organized as follows. We first derive $\like(\lambda|\vecw(\vectheta))$ assuming identical weights in Sec.~\ref{sec:constructing}, then extend it to arbitrary weights in Sec.~\ref{sec:extending}. With this in hand, we calculate an analytic expression for Eq.~\eqref{eq:generalpoisson} using Eq.~\eqref{eq:posterior} under a uniform $\prob(\lambda)$ in Sec.~\ref{sec:effective}. In Sec.~\ref{sec:priors} we briefly discuss a family of distributions as possible alternative priors. In Sec.~\ref{sec:llhconvergence} we show that our effective likelihood converges to Eq.~\eqref{eq:mcpoisson} in the limit of large MC size. Finally, in Sec.~\ref{sec:llhbehavior} we provide some intuition on the behavior of our generalized likelihood. Equation \eqref{eq:parametrizedpoisson}, along with the definitions of $\mu$ and $\sigma^2$ given in Eq.~\eqref{eq:musigma}, constitute the primary result of this work.

\subsection{Derivation of $\like (\lambda|\vecw(\vectheta))$ for identical weights\label{sec:constructing}}
In this section we derive $\like(\lambda|\vecw(\vectheta))$ for identical weights. We will show that $\like(\lambda|\vecw(\vectheta))$ can be written in terms of two quantities
\begin{equation}\label{eq:musigma}
\mu \equiv \sum_{i=1}^m w_i~\textmd{and}~\sigma^2 \equiv \sum_{i=1}^m w_i^2
\end{equation}
for a bin with $m$ MC events.

For identical weights, $w \equiv w_i~\forall i$, the following equalities hold:
\begin{equation}\label{eq:ids}
\mu=wm \textmd{,}~ \sigma^2=w^2 m \textmd{,}~ w=\sigma^2/\mu \textmd{, and }~ m = \mu^2/\sigma^2.
\end{equation} 
Assume that $m$ is the outcome of sampling a Poisson-distributed random variable $M$ with probability mass function
\begin{equation}\label{eq:mcprob}
\mathrm{Poisson}(M=m;\bar m) = \frac{e^{-\bar m} {\bar m}^m}{m!},
\end{equation}
where $\bar m$ is the mean of the distribution. Further, assume that the expected number of data events $\lambda=w \bar m$ so that $\bar m = \lambda/w$. Substituting back into Eq.~\eqref{eq:mcprob}, we can interpret $\mathrm{Poisson}(M=m;\bar m)$ as a likelihood function of $\lambda$
\begin{equation}
\like(\lambda|\vecw(\vectheta))=\like(\lambda|\mu, \sigma)=\frac{e^{-\lambda\mu/\sigma^2}\left(\lambda\mu/\sigma^2\right)^{\mu^2/\sigma^2}}{(\mu^2/\sigma^2)!},
\label{eq:poisson_conditional}
\end{equation}
as $\mu$ and $\sigma$ fully specify $\vecw(\vectheta)$ for identical weights. 

\subsection{Extension to arbitrary weights\label{sec:extending}}
The derivation above assumed identical weights. For arbitrary weights, $\mu$ is an outcome sampled from a compound Poisson distribution (CPD), which can be approximated by a scaled Poisson distribution (SPD) by matching the first and second moments of the two distributions~\cite{Bohm:2013gla}. In order to make the connection, first rewrite $\mu$ and $\sigma^2$ as
\begin{equation}\label{eq:effparameters}
\mu= \weff \meff~\textmd{and}~\sigma^2 = \weff^2 \meff,
\end{equation}
where $\meff$ is the effective number of MC events and $\weff$ the effective weight. From Eq.~\eqref{eq:ids} these are given by: $\meff = \mu^2/\sigma^2$ and $\weff=\sigma^2/\mu$. Next, assume $\bar m = \lambda/\weff$ and
\begin{align}
\label{eq:probmeff}
\like(\bar m|\meff)&= \frac{e^{-\bar m}{\bar m}^{\meff}}{\Gamma(\meff+1)},
\end{align}
where $\lambda$ again is the expected number of events in data. Equation \eqref{eq:probmeff} can be written as a likelihood function of $\lambda$,
\begin{equation}
\like(\lambda|\vecw(\vectheta))=\like(\lambda|\mu, \sigma)=\frac{e^{-\lambda\mu/\sigma^2}\left(\lambda\mu/\sigma^2\right)^{\mu^2/\sigma^2}}{\Gamma(\mu^2/\sigma^2+1)},
\label{eq:poisson_conditional_arb}
\end{equation}
which is identical to Eq.~\eqref{eq:poisson_conditional} except the denominator is now a gamma function instead of a factorial. However, since the denominator does not depend on $\lambda$ it cancels out in Eq.~\eqref{eq:posterior}.

To understand this approximation, note that the maximum likelihood in Eq.~\eqref{eq:probmeff} occurs when $\bar m = \meff$. The first and second moments of the SPD random variable $\weff M$, where $M \sim \mathrm{Poisson}(\meff)$, are given by
\begin{align}
\mathrm{E}[\weff M] &= \weff \meff \\
&= \mu, \nonumber
\end{align}
and 
\begin{align}
\mathrm{Var}[\weff M] &= \weff^2 \meff \\
&= \sigma^2. \nonumber
\end{align}
This shows that the SPD, under the maximum likelihood solution for the given MC realization, has first and second moments that match the sample mean, $\mu$, and variance, $\sigma^2$, respectively. These are equal to the first and second moments of the CPD as described in~\cite{Bohm:2013gla}. By assuming that $\mu$ is drawn from a SPD, we can treat $\mu$ and $\sigma$ as outcomes that fix the likelihood function of the underlying scaled expectation $\lambda$, analogous to the case of identical weights. Because both the first and second moments are matched, this approximation accounts for the variance of the CPD unlike $\lbarlow$, which only accounts for the mean. Thus, while $\lbarlow$ is valid only for the case of narrow weight distributions, our approximation remains valid for broader distributions.

\subsection{The effective likelihood\label{sec:effective}}
Now that we have an expression for $\like(\lambda|\vecw(\vectheta))$ from the MC, we can proceed to compute Eq.~\eqref{eq:generalpoisson} under a uniform $\prob(\lambda)$. To simplify the notation, let
\begin{equation}\label{eq:alphabetamc}
\agpar = \frac{\mu^2}{\sigma^2}+1~\textmd{and}~\bgpar=\frac{\mu}{\sigma^2}.
\end{equation}
Then, assuming a uniform $\prob(\lambda)$ and substituting Eq.~\eqref{eq:poisson_conditional_arb} for $\like(\lambda|\vecw(\vectheta))$ in Eq.~\eqref{eq:posterior} we obtain
\begin{align} \label{eq:theposterior}
\prob(\lambda|\vecw(\vectheta)) &= \bgpar \frac{ e^{-\lambda \bgpar}(\lambda \bgpar )^{\agpar-1}}{\Gamma(\agpar)}\nonumber \\
&= \frac{e^{-\lambda \bgpar } \lambda^{\agpar-1} \bgpar^{\agpar}}{\Gamma(\agpar)} \nonumber \\
&= \gprob(\lambda;\agpar, \bgpar),
\end{align}
where $\Gamma$ is the gamma function and $\gprob$ the gamma distribution with shape parameter $\agpar$ and inverse-scale parameter $\bgpar$.
Note that in going from Eq.~\eqref{eq:poisson_conditional_arb} to Eq.~\eqref{eq:theposterior} $\mu$ and $\sigma^2$ go from random variates for a particular $\lambda$ to parameters that govern the probability density of $\lambda$. With this choice of $\like(\lambda|\vecw(\vectheta))$ and $\prob(\lambda)$, we can rewrite $\like_{\rm General}$ from Eq.~\eqref{eq:generalpoisson} as
\begin{align}
\mcl(\vectheta|k) &=\int_{0}^{\infty}~\frac{\lambda^k e^{-\lambda}}{k!}\gprob(\lambda;\agpar, \bgpar)~d\lambda \\
&= \frac{\bgpar^\agpar\Gamma\left(k+\agpar\right)}{k!\left(1+\bgpar\right)^{k+\agpar}\Gamma\left(\agpar\right)} \\
&= \left(\frac{\mu}{\sigma^2}\right)^{\frac{\mu^2}{\sigma^2}+1}\Gamma\left(k+\frac{\mu^2}{\sigma^2}+1\right)\left[k!\left(1+\frac{\mu}{\sigma^2}\right)^{k+\frac{\mu^2}{\sigma^2}+1}\Gamma\left(\frac{\mu^2}{\sigma^2}+1\right)\right]^{-1}, \label{eq:parametrizedpoisson}
\end{align}
where $\mu$ and $\sigma^2$ depend on $\vectheta$ through $\vecw$.

\subsection{A family of likelihoods\label{sec:priors}}

It is possible to generalize the choice of $\alpha$ and $\beta$ in Eq.~\eqref{eq:alphabetamc} by choosing a particular form of $\prob(\lambda)$. Since the distribution of interest is a Poisson distribution, a well-motivated choice of $\prob(\lambda)$ is a gamma distribution (the conjugate prior of the Poisson distribution)~\cite{Fink97acompendium}; also see~\cite{Glusenkamp:2017rlp} for a recent discussion. Thus we set $\prob(\lambda) = \gprob(\lambda; a, b)$, where $a$ and $b$ are the shape and inverse-scale parameters of the gamma distribution, respectively. These hyper-parameters dictate the distribution of the Poisson parameter $\lambda$~\cite{bernardo2009bayesian}. In line with our previous discussion, the gamma distribution prior implies that Eq.~\eqref{eq:alphabetamc} becomes
\begin{equation}\label{eq:alphabetagamma}
\agpar = \frac{\mu^2}{\sigma^2}+a~\textmd{and}~\bgpar=\frac{\mu}{\sigma^2}+b.
\end{equation}
The rest of the likelihood derivation remains the same. This allows the choice of specific values for $a$ and $b$ to satisfy certain properties. Equation~\eqref{eq:alphabetamc} is obtained with $a=1$ and $b=0$, corresponding to the uniform prior discussed above. Another interesting choice is to require that the mean and variance of $\prob(\lambda|\mu, \sigma)$ match $\mu$ and $\sigma^2$, respectively. This can be achieved by setting $a=b=0$, and we refer to this parameter assignment as $\meanl$. In the case of identical weights, $\meanl$ is equivalent to Eq.~(20) in~\cite{Glusenkamp:2017rlp}. Both choices are improper priors, as technically they are limiting cases of the gamma distribution. However, we can use them to obtain proper $\prob(\lambda|\mu, \sigma)$ distributions.

In~\cite{Glusenkamp:2017rlp}, a convolutional approach is suggested for handling arbitrary weights. We refer to this likelihood as $\gl$. Each weighted MC event has $\prob(\lambda_i|w_i)=\gprob(\lambda_i; 1, 1/w_i)$, corresponding to the prior $\prob(\lambda_i) = \gprob(\lambda_i;0,0)$, such that $\lambda = \sum_i^m \lambda_i$. The likelihood $\meanl$ is a good analytic approximation of the more computationally expensive calculation given in~\cite{Glusenkamp:2017rlp} for $\gl$. The latter has time complexity $\mathcal{O}(k^2 m)$ where $k$ and $m$ are the number of data and MC events in the bin respectively. When assuming uniform priors, the convolutional approach does not recover Eq.~\eqref{eq:theposterior} for identical weights, so it cannot be used as a generalization of $\mcl$.

\subsection{Convergence of the effective likelihood\label{sec:llhconvergence}}

In this section we will show that, if the relative uncertainty of the bin content vanishes as MC size increases, $\mcl$ and $\meanl$ both converge to $\adhoc$. 

For positive weights $w_i$, the relative uncertainty $\sigma/\mu$ is bounded between zero and one. Uncertainty as large as the estimated quantity, $\sigma/\mu=1$, occurs if and only if $m=1$. In the limit that $\sigma/\mu$ goes to zero, Eq.~\eqref{eq:theposterior} converges to $\delta(\lambda - \mu)$ and $\mcl$ and $\meanl$ both go to $\adhoc$. We can see this by noting that the shape parameter, $\agpar$, goes to infinity as the MC relative uncertainty goes to zero, turning the gamma distribution into a Gaussian distribution of mean $\agpar / \bgpar$ and variance $\agpar / \bgpar^2$. This Gaussian converges to $\delta(\lambda-\mu)$ in the limit of vanishing $\sigma/\mu$. Substituting into Eq.~\eqref{eq:generalpoisson}, we recover Eq.~\eqref{eq:mcpoisson}, which converges to Eq.~\eqref{eq:poisson} in the large MC limit.

It remains to be shown that the relative uncertainty of the bin content vanishes as MC size increases. For identical weights,
\begin{equation}
\lim_{m\to\infty} \frac{\sigma_{\rm identical}}{\mu_{\rm identical}} = \lim_{m\to\infty} \frac{1}{\sqrt{m}} = 0 .
\end{equation}
For arbitrary weights, the limit can be written in terms of the running average of $w_i$ and $w_i^2$ as
\begin{align}
\lim_{m\to\infty} \frac{\sigma}{\mu} &= \lim_{m\to\infty} \frac{\sqrt{\langle w^2 \rangle_m}}{\langle w\rangle_m\sqrt{m}},
\end{align}
where $\langle w \rangle_m$ is the average over $w_i$ and $\langle w^2 \rangle_m$ the average over $w_i^2$ for $i \leq m$. This shows that as long as $\langle w^2 \rangle_m$ does not grow much faster than $\langle w \rangle_m^2$, the limit will converge to zero. For weight distributions with positive support and finite, non-zero mean, this should be the case.

\subsection{Behavior of the effective likelihood\label{sec:llhbehavior}}

\begin{figure}[htp]
\centering
\includegraphics[width=0.5856\linewidth]{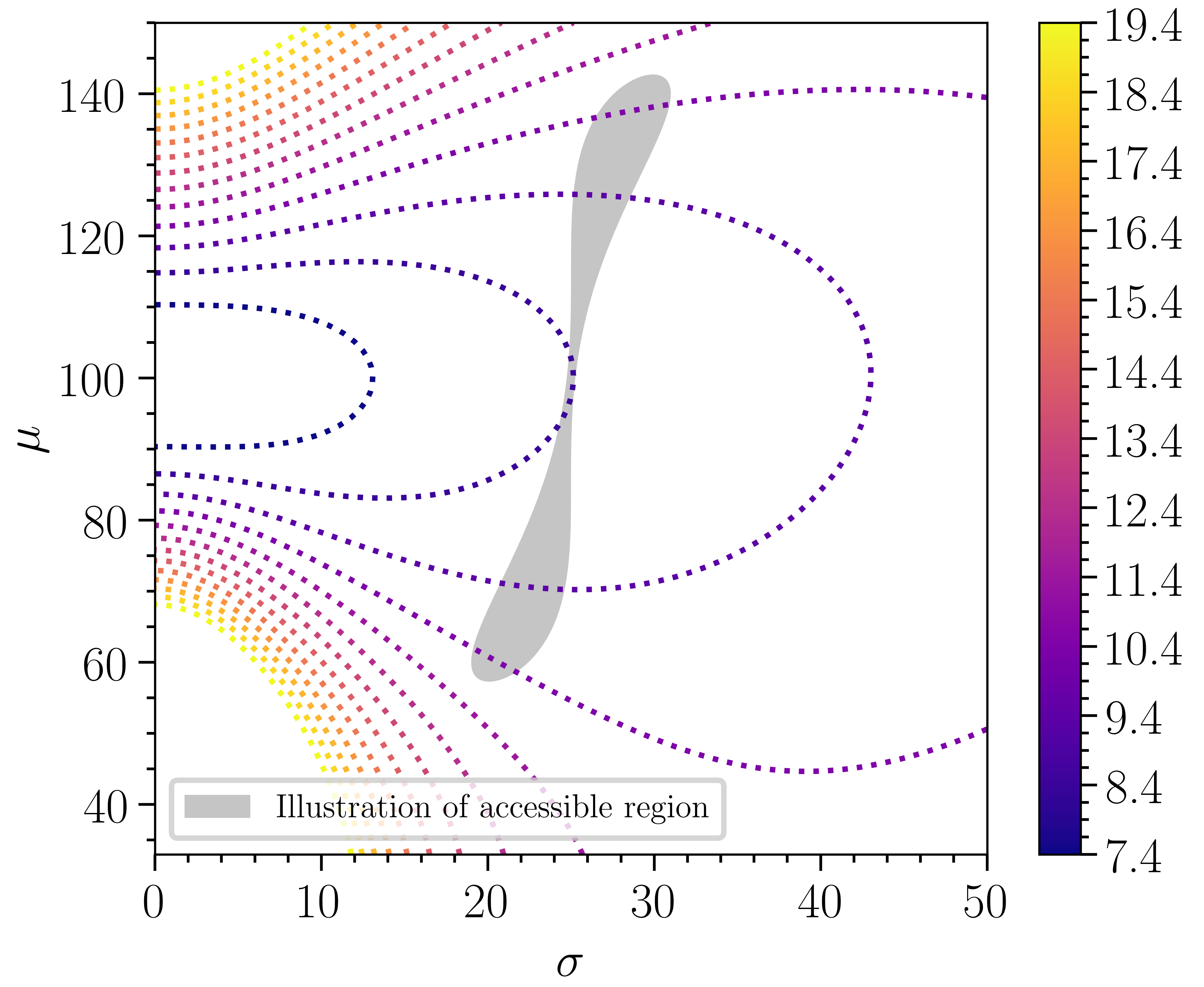}
\caption{\textbf{\textit{Likelihood contours and accessible region.}} Contours of constant $\lmc(\mu,\sigma|k=100)$. The accessible region (gray) illustrates where values of $\mu$ and $\sigma$ may lie for a hypothetical physics model. A minimization over $\vectheta$ can be thought of as a constrained minimization over the accessible region in $\mu$ and $\sigma$. Note that as $\sigma$ increases the contours broaden.}
\label{fig:contour}
\end{figure}

It is instructive to examine the behavior of $\mcl$ for a single bin. It is standard to work with the log-likelihood $l(\mu,\sigma|k) \equiv -2\ln \like(\mu, \sigma|k)$ and we do so here. Figure \ref{fig:contour} shows the contour lines for $\lmc(\mu,\sigma|k=100)$. Since $\mu$ and $\sigma$ are both dependent on the same underlying parameters, $\vectheta$, a minimization over $\vectheta$ can be thought of as a constrained minimization over $\mu$ and $\sigma$. This is visualized as the gray region in Fig.~\ref{fig:contour}, which indicates where $\mu$ and $\sigma$ are allowed to vary for some physics model\footnote{A general bound for positive weights is $\sigma \leq \mu \leq \sigma \sqrt{m}$ which can be seen from their definitions.}. Similarly, we can also visualize the standard Poisson log-likelihood, $\lpoisson(\mu|k=100)$, which is simply $\lmc$ constrained along the line $\sigma=0$. %In the case of multiple bins, the likelihood is a superposition of the likelihood in each bin.

\begin{figure}[htp]
\centering
\centering
    \subfloat{
        \includegraphics[width=0.48\linewidth]{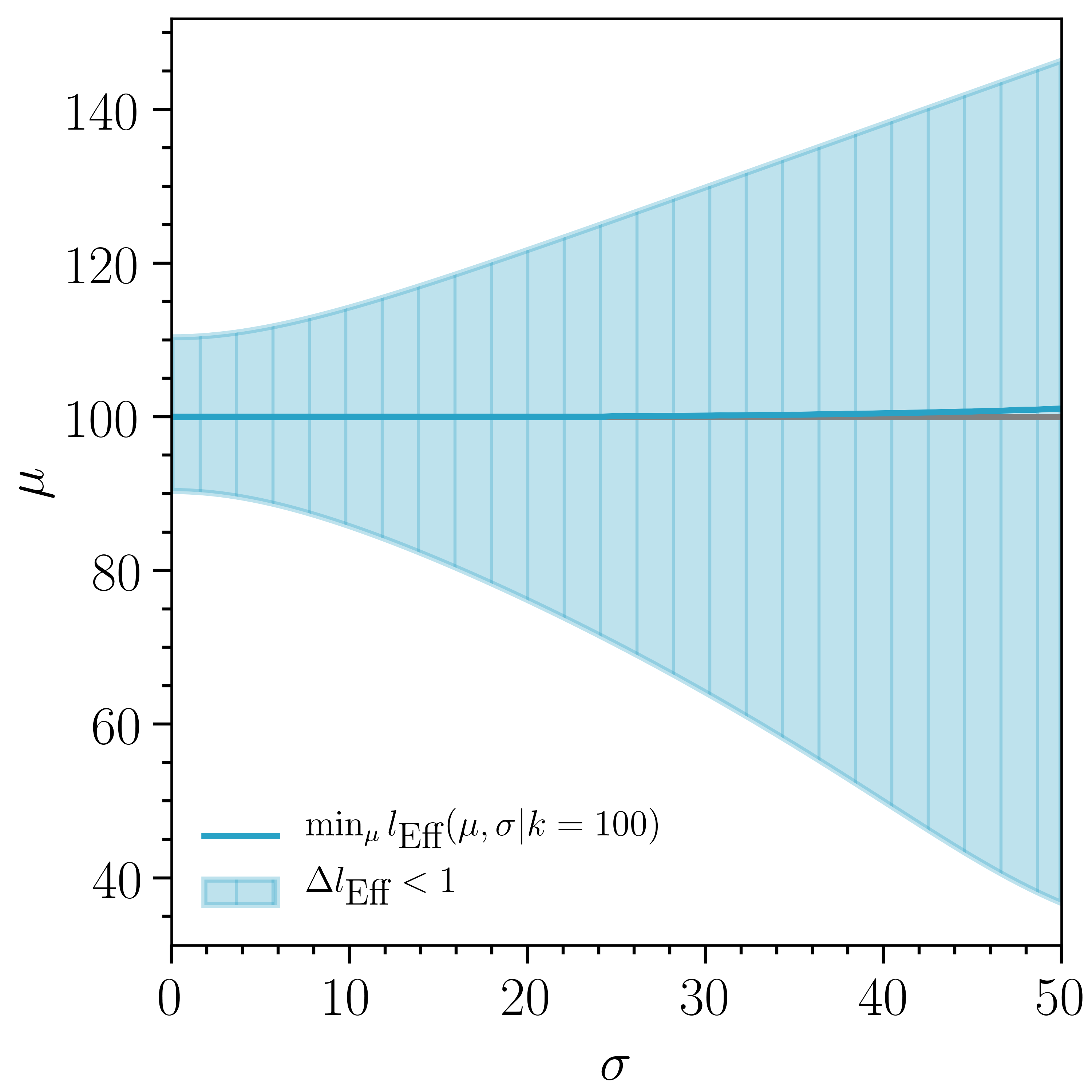}
    }
    \subfloat{
        \includegraphics[width=0.48\linewidth]{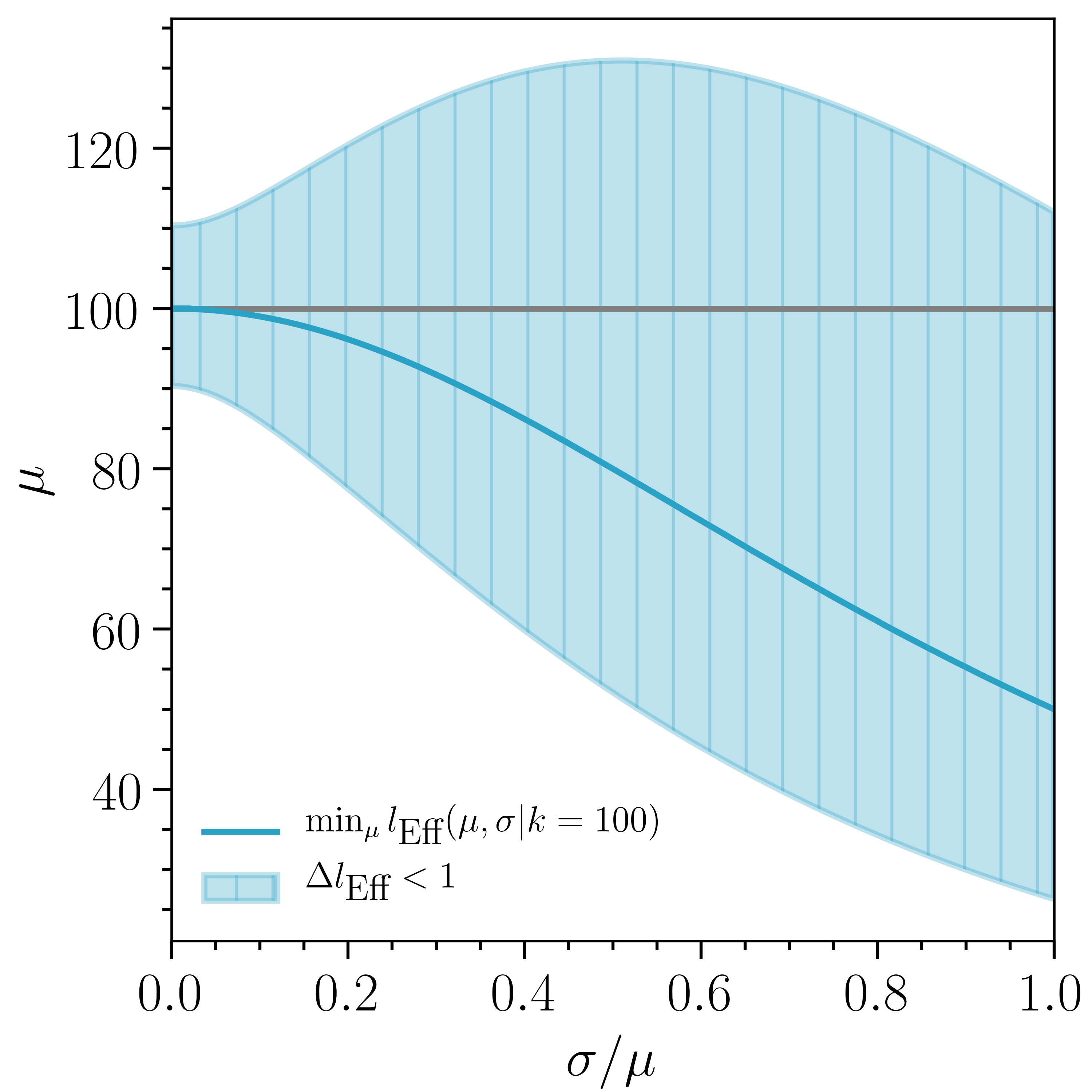}
    }
\caption{\textbf{\textit{Slices of $\lmc$ for two accessible regions.}} This figure shows $\lmc(\mu, \sigma|k=100)$ minimized over $\mu$ while $\sigma$ (left) and $\sigma/\mu$ (right) are held fixed. The minimum, $\hatmc$, is shown as the solid blue line running through the center of the shaded regions. The shaded regions indicate where $\lmc(\mu, \sigma|k) - \lmc(\hatmc, \sigma|k) < 1$. As $\sigma$ goes to zero, the Poisson best-fit $\hatpoisson=100$ is obtained. For fixed $\sigma/\mu$, $\hatmc$ deviates from $\hatpoisson$ as $\sigma/\mu$ increases.}
\label{fig:llhmin}
\end{figure}

To further illustrate the effect of the accessible region, we minimize $\lmc$ over $\mu$ for two possible constraints: fixed $\sigma$ and fixed $\sigma/\mu$. In terms of Eq.~\eqref{eq:musigma}, a sufficient but not necessary condition for constant $\sigma/\mu$ with varying $\mu$ is equal weights, and a necessary but not sufficient condition for constant $\sigma$ with varying $\mu$ is $m \geq 2$. For a standard Poisson likelihood, $\hatpoisson\equiv \min_\mu\lpoisson(\mu|k) = k$. Figure \ref{fig:llhmin} shows $\hatmc \equiv \min_\mu \lmc(\mu,\sigma|k=100)$ as well as the region where $\lmc(\mu, \sigma|k) - \lmc(\hatmu, \sigma|k) < 1$ for fixed $\sigma$ (left) and fixed $\sigma/\mu$ (right). Note that the shaded regions for fixed $\sigma$ are calculated without requiring that $\mu \geq \sigma$, which would be the case for Eq.~\eqref{eq:musigma}. As $\sigma$ goes to zero, the Poisson best-fit and Wilks' $1\sigma$ interval are recovered. As $\sigma$ or $\sigma/\mu$ increases, the shaded region becomes wider, as expected. For fixed $\sigma$, $\hatmc$ does not deviate much from $\hatpoisson$, while for fixed $\sigma/\mu$, $\hatmc$ deviates from $\hatpoisson$ as $\sigma/\mu$ increases. The shaded regions correspond to the $1\sigma$ interval assuming the approximation from Wilks' theorem and give a sense of the shape of $\mcl$ projected onto one-dimensional slices.

\section{Example and performance}\label{sec:example}

In practice, likelihoods such as those discussed above are used to estimate physical parameters from data. As discussed in Sec.~\ref{sec:intro}, weighted MC is often used to compute the likelihood of a particular physical scenario given the observed data. Statements are then made about the physical scenarios either by maximizing the likelihood or by examining the posterior distribution assuming some priors. We examine a toy experiment where we measure the mode, $\Omega$, and normalization, $\Phi$, of a Gaussian-distributed signal against a steeply falling inverse power-law background. The performance of $\mcl$ is evaluated and compared against other likelihoods.

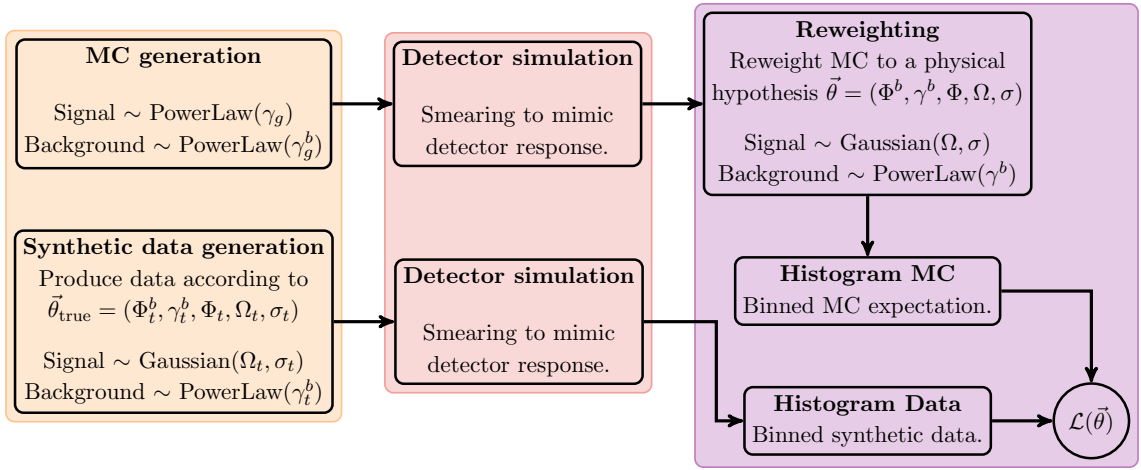
\begin{figure}[htp]
\resizebox{\textwidth}{!}{
\begin{tikzpicture}[node distance=1cm, auto,]

%nodes
\node[solidbox] (mc_generation) {\centering {\bf MC generation} \\
	\begin{center}
		Signal $\sim$ PowerLaw($\gamma_g$) \\
		Background  $\sim$ PowerLaw($\gamma_g^b$)
	\end{center}
};

\node[solidbox, right=of mc_generation] (detector_simulation) {\centering {\bf Detector simulation} \\
    \begin{center}
	Smearing to mimic\\ detector response.
	\end{center}
};

\node[solidbox,right=of detector_simulation] (reweighting) {\centering {\bf Reweighting} \\
	Reweight MC to a physical\\  hypothesis $\vec\theta = (\Phi^{b},\gamma^b,\Phi,\Omega,\sigma)$
	\begin{center}
		Signal $\sim$ Gaussian($\Omega,\sigma$) \\
		Background $\sim$ PowerLaw($\gamma^b$)
	\end{center}
};

\node[solidbox,below=of reweighting] (mc_histogram) {\centering {\bf Histogram MC} \\
	Binned MC expectation.
};

\node[solidbox, below=of mc_generation] (data_generation) {\centering {\bf Synthetic data generation} \\
	Produce data according to\\ $\vec\theta_{\mathrm{true}} = (\Phi_t^{b},\gamma_t^b,\Phi_t,\Omega_t,\sigma_t)$
	\begin{center}
		Signal $\sim$ Gaussian($\Omega_t,\sigma_t$) \\
		Background $\sim$ PowerLaw($\gamma_t^b$)
	\end{center}
};

\node[solidbox, right=of data_generation] (data_detector_simulation) {\centering {\bf Detector simulation} \\
    \begin{center}
	Smearing to mimic\\ detector response.
	\end{center}
};

\node[solidbox,below=of mc_histogram] (data_histogram) {\centering {\bf Histogram Data} \\
	Binned synthetic data.
};

\node[circle, minimum size=1cm, color=black, draw=black, very thick, right=of data_histogram] (likelihood) {\centering {\bf $\mathcal{L}(\vec\theta)$}
};

% arrows

\draw [->,line width=1.5pt] (mc_generation) -- (detector_simulation);
\draw [->,line width=1.5pt] (detector_simulation) -- (reweighting);
\draw [->,line width=1.5pt] (reweighting) -- (mc_histogram);
\draw [->,line width=1.5pt] (mc_histogram) -| (likelihood);

\draw [->,line width=1.5pt] (data_generation) -- (data_detector_simulation);
\draw [->,line width=1.5pt] (data_detector_simulation.east) |- ($(data_detector_simulation.east) + (1.2,0.0)$) |- (data_histogram.west);
\draw [->,line width=1.5pt] (data_histogram) -- (likelihood);

% boxes

\begin{pgfonlayer}{background}
  \node[bigboxGeneration] [fit = (mc_generation) (data_generation)] {};
\end{pgfonlayer}

\begin{pgfonlayer}{background}
  \node[bigboxDetector] [fit = (detector_simulation) (data_detector_simulation)] {};
\end{pgfonlayer}

\begin{pgfonlayer}{background}
  \node[bigboxAnalysis] [fit = (reweighting) (mc_histogram) (data_histogram) (likelihood)] {};
\end{pgfonlayer}
\end{tikzpicture}
}% end resize box
\caption{\textbf{\textit{Diagram of toy experiment steps.}} The three colored boxes indicate the three steps of our toy experiment. The left box (almond) summarizes the MC and data generation. The center box (salmon) indicates the step in which we apply the detector response. The right box (lilac) summarizes the MC reweighting, data and MC histogramming, and final likelihood evaluation from the histograms. This final lilac box is repeated for each likelihood evaluation.}
\label{fig:mc_diagram}
\end{figure}

For our toy experiment, we generate the true energies, $E_t$, of synthetic data events from a background falling as $(E_t/100 \mathrm{GeV})^{-\gamma_t^b}$, where $\gamma_t^b=3.07$, and a Gaussian signal centered at $\Omega_t=125$~GeV with width of $\sigma_t=2$~GeV and normalization $\Phi_t=5013$ for a fixed number of expected events. Our imaginary detector is sensitive in the 100--160~GeV range. To simulate the effect of a real detector, the true energy, $E_t$, is smeared by 5\% for background and 3\% for signal to obtain event-by-event reconstructed energies, $E_r$. We generate a total number of MC events, $N_{\mathrm{MC}}$, split evenly between the components. Generation is performed assuming inverse power-law distributions of $(E_t/100 \mathrm{GeV})^{-\gamma_g}$ for signal and $(E_t/100 \mathrm{GeV})^{-\gamma_g^b}$ for background. We choose $\gamma_g=1$ and $\gamma_g^b=2$. Reweighting of the MC can then be performed as a function of $E_t$ and forward-folded onto distributions in $E_r$ over which the events are histogrammed and likelihoods evaluated. A diagram of the steps described above is shown in Fig.~\ref{fig:mc_diagram}. For all toy experiments, the background component, $(\Phi^b,\gamma^b)$, and the signal width, $\sigma$, are kept fixed to their true values. Only the signal mean, $\Omega$, and normalization, $\Phi$, are treated as free parameters.

\subsection{Point estimation}
\label{sec:pointestimation}

\begin{figure}[htp]
\centering
    \includegraphics[width=1.\linewidth]{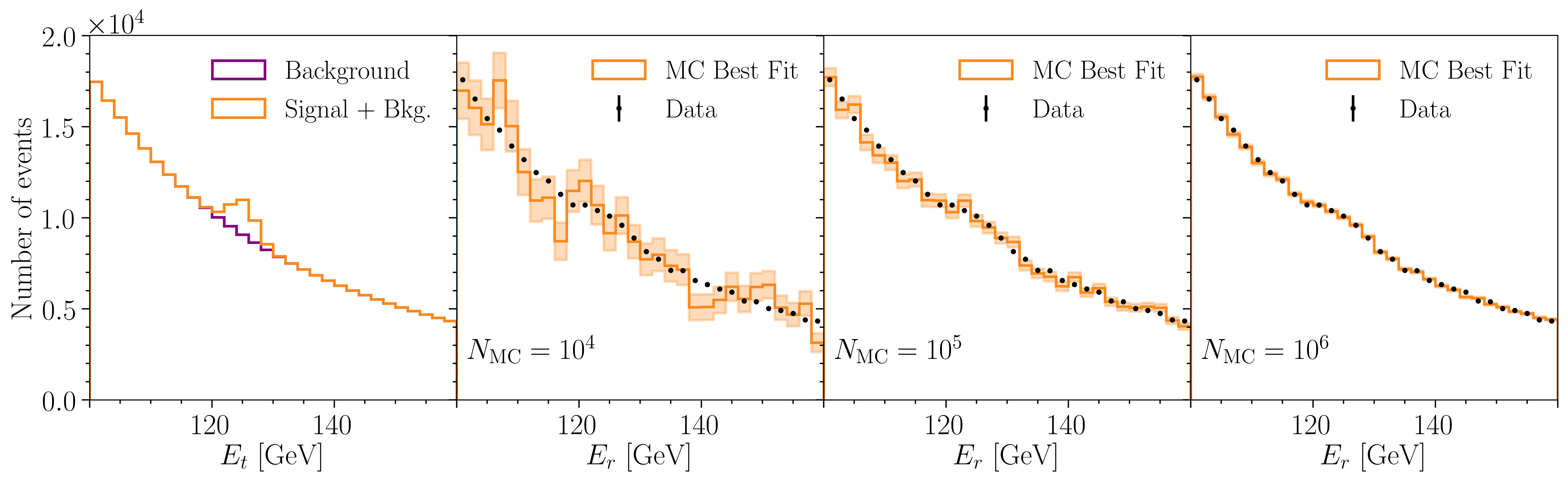}
\caption{\textbf{\textit{Benchmark scenario.}} Leftmost panel: the underlying distribution that data is drawn from with background (purple) and total rate (orange). Right three panels: observed data (black), $\mcl$ best-fit MC distributions (orange), and MC uncertainties (orange band), with increasing MC size from left to right. The background component and signal width are fixed, while the mean and normalization of the signal peak are fit by maximizing the likelihood with respect to those parameters.}
\label{fig:toymc}
\end{figure}

Figure~\ref{fig:toymc} shows the expectation in $E_t$ as well as the data and $\mcl$ best-fit distributions in $E_r$. The leftmost panel shows the expectation for both signal and background assuming no smearing in $E_t$. The three other panels show the smeared, $E_r$, distribution for data (black) and the best-fit result from $\mcl$ for three different MC datasets (orange) of varying MC size. The smeared shape of the signal peak is clearly visible in data, but not in the smallest size MC. As the MC increases in size, the best-fit MC can be seen to converge to data.

\begin{table}[htp]
\centering
\begin{tabular}{l r r r}
\toprule
Likelihood & $N_\mathrm{MC}=10^4$ & $10^5$ & $10^6$ \\
\midrule
%\hhline{ | = # = # = # = # = # = # = # = # = # = # = | }
$\adhoc$ & $(127.0,6368.0)$ & $(124.7,5655.7)$ & $(125.1,4888.5)$ \\ \hline
$\mcl$ & $(127.1,6077.1)$ & $(124.7,5576.0)$ & $(125.1,4889.4)$ \\
\bottomrule
\end{tabular}
\caption{\textbf{\textit{Best-fit parameters}} For the toy experiment shown in Fig.~\ref{fig:toymc}, best-fit parameters using $\adhoc$ and $\mcl$ are shown. The columns in the table are for the different MC sizes. The two numbers in parenthesis in each entry correspond to $\Omega$ and $\Phi$, respectively.}
\label{tbl:pointestimator}
\end{table}

The best-fit values for the example shown in Fig.~\ref{fig:toymc} are given in Table~\ref{tbl:pointestimator} for $\mcl$ and $\adhoc$. As point estimators, both likelihoods return similar values. This is driven by the fact that the same underlying MC distribution is used to fit to the data. The effect of convoluting $\prob(\lambda|\vecw(\vectheta))$ mostly serves to broaden the likelihood space, while preserving the maximum within the constraints described in Sec.~\ref{sec:llhbehavior}. In the large MC limit, both likelihoods can be used for unbiased point estimation, provided that the likelihood space is smooth enough for standard minimization techniques to probe the global minimum.

\subsection{Coverage}
\label{sec:coverage}
Due to the higher computational cost of computing frequentist confidence intervals by generating pseudodata to estimate the test-statistic ($\mathcal{TS}$) distribution, it is common to use the approximation given by Wilks' theorem for the cases where the underlying hypotheses hold. In the case of small MC, a likelihood description that neglects MC uncertainties may lead to undercoverage even for a large data sample. In this section, we will use  $\mathcal{TS}=\Delta l=l(\vectheta_{\rm true})-l(\hat{\vectheta})$, where $\vectheta_{\rm true}$ and $\hat{\vectheta}$ correspond to the true and best-fit $(\Omega, \Phi)$, respectively. We evaluate the coverage properties, computed using the asymptotic approximation given by Wilks' theorem, of the two-dimensional fit over $(\Omega, \Phi)$ for several likelihood constructions. These include the modified-$\chi^2$, $\adhoc$, $\lbarlow$, $\meanl$, and $\mcl$. These five test-statistics were chosen on the basis of their computation speed and as tests of different approaches towards the treatment of weighted MC. Note that using Wilks' theorem is an approximation and in general we encourage the reader to perform coverage tests for their own particular setup.

\begin{figure}[htp]
\centering
\centering
    \subfloat{
        \includegraphics[width=0.48\linewidth]{{{fig/fig5_1_1e3}}}
    }
    \subfloat{
        \includegraphics[width=0.48\linewidth]{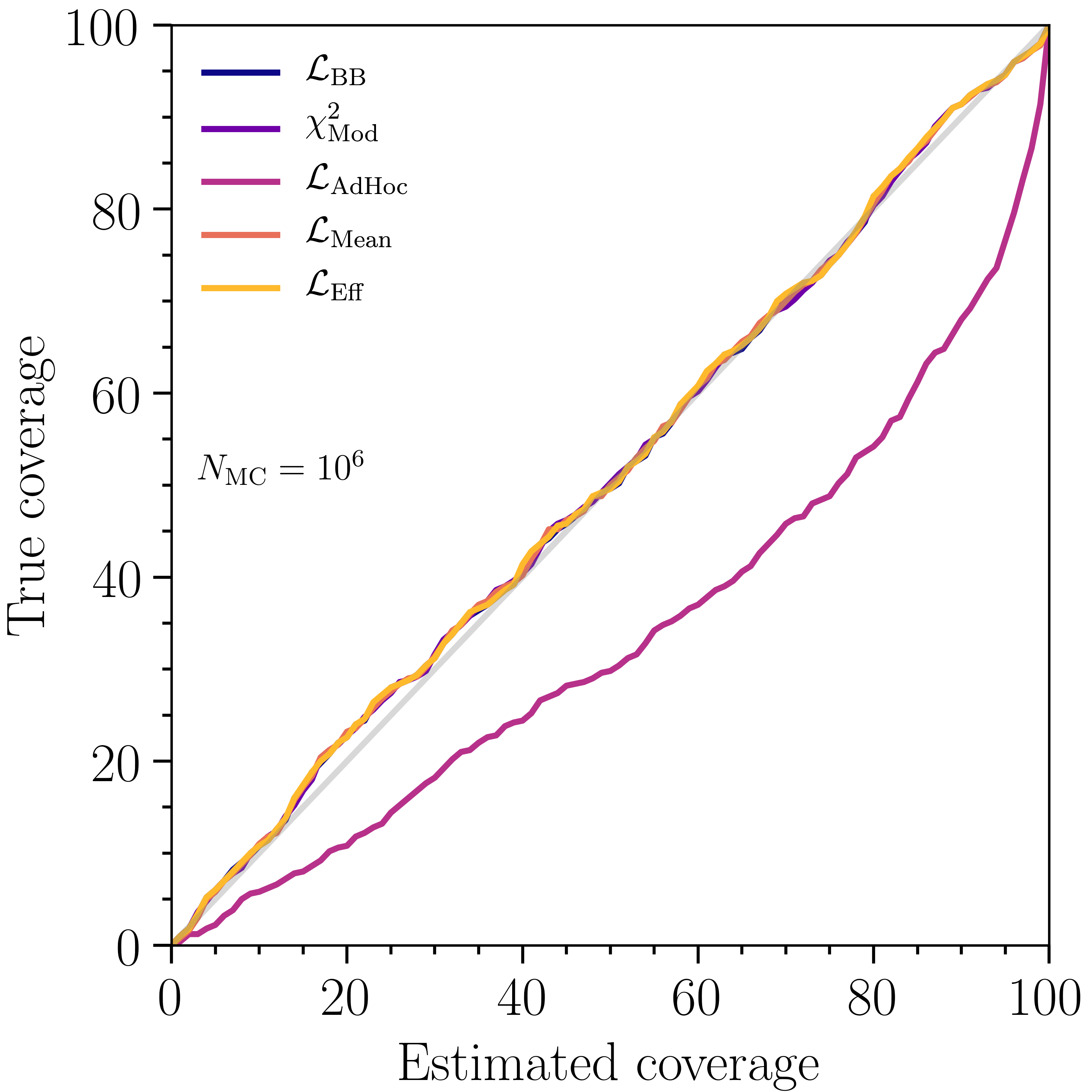}
    }
\caption{\textbf{\textit{Coverage properties.}} True coverage of the Wilks' confidence interval for two sets of toy experiments with different MC sizes: $10^3$ events (left) and $10^6$ events (right). The ad hoc Poisson likelihood severely undercovers. The modified-$\chi^2$, $\lbarlow$, and $\meanl$ also undercover for small MC size. The effective likelihood, $\mcl$, derived in Sec.~\ref{sec:constructing} performs best.}
\label{fig:coverage}
\end{figure}

Several configurations were tested, all under the assumptions of the toy experiment described in Sec.~\ref{sec:pointestimation}. The MC was generated for two different settings of the total number of events: $10^3$ and $10^6$. For each setting, 500 toy experiments were generated, their best-fits found, and their $\Delta l$ evaluated. Each toy experiment was classified as covering $\vectheta_{\rm true}$ at a specified level $p$ if $\Delta l < I(p;2)$, where $I$ is the inverse of the $\chi^2$ cumulative density function and $2$ indicates the number of degrees of freedom.

Figure~\ref{fig:coverage} shows the percentage of times the true parameters were within the confidence intervals at level $p$ as a function of the estimated coverage percentile for that level. First note that, as expected, the true coverage is highly dependent on MC size, with higher MC size leading towards improved agreement. In the case of $N_\mathrm{MC}=10^3$, $\lbarlow$, $\meanl$, modified-$\chi^2$, and $\adhoc$ all undercover to varying degrees of severeness. For $N_\mathrm{MC}=10^6$, $\adhoc$ still undercovers, which is not surprising as it presumes zero MC uncertainty, but the other likelihoods exhibit good agreement. In this benchmark test, $\mcl$ exhibits the best coverage properties. However, note that using Wilks’ theorem in order to evaluate confidence intervals implies an asymptotic approximation. In general, this approximation does not necessarily have to hold and we encourage the reader to always perform their own coverage tests suitable for their particular experimental setup.

\subsection{Posterior distributions}
\label{sec:posterior}

\begin{figure}[ht]
\centering
    \includegraphics[width=1\linewidth]{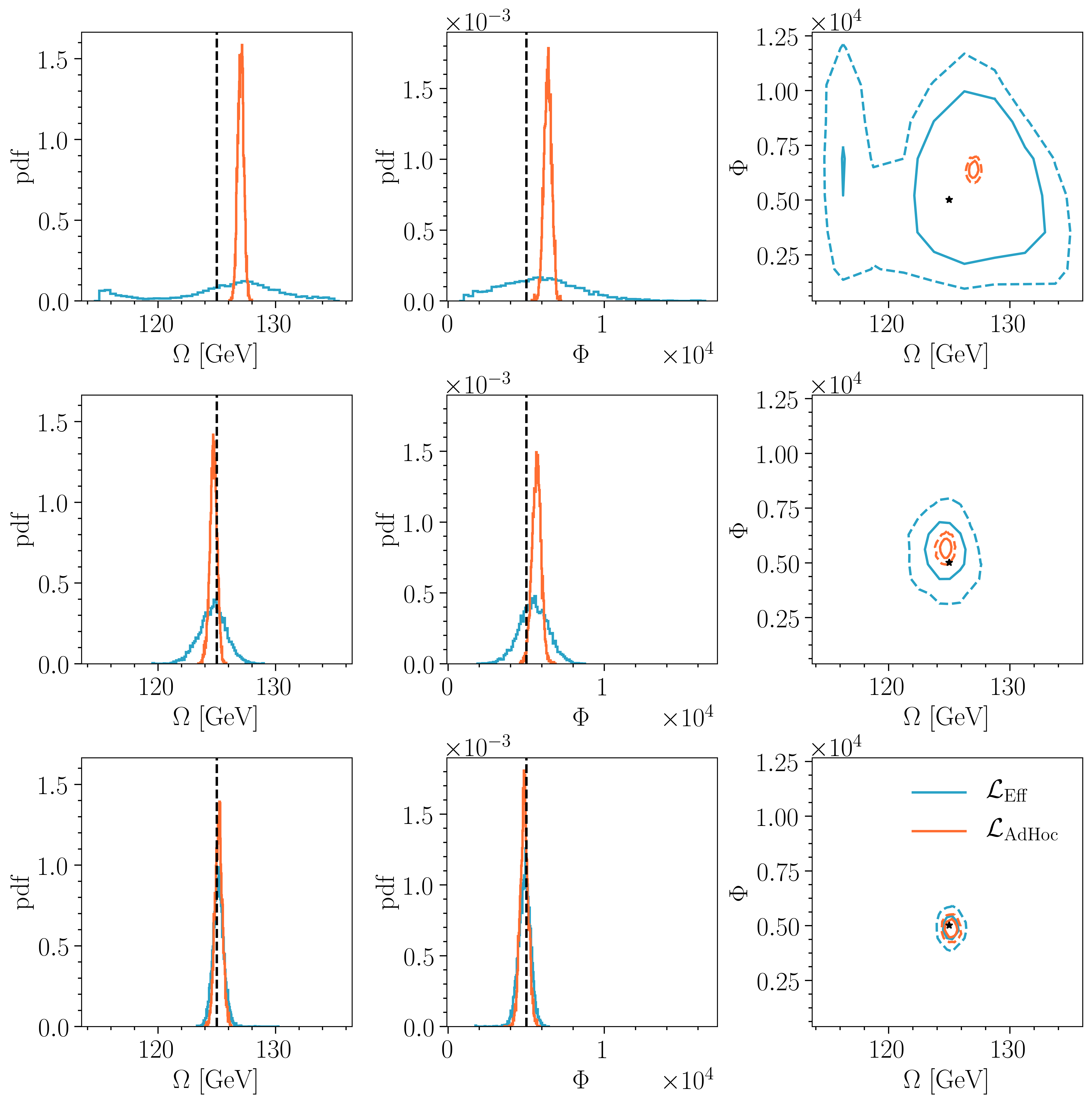}
\caption{\textbf{\textit{Posterior distributions in parameter space.}} Comparison of $\prob(\vectheta|k)$ for $\mcl$ (blue) and $\adhoc$ (orange). Each horizontal row above uses a different MC set size, with $N_\mathrm{MC}=10^4$, $10^5$, and $10^6$ from top to bottom. The left and center column show the marginal posterior distribution for the mass, $\Omega$, and normalization, $\Phi$, respectively. The true value is indicated by the dashed, vertical line. The rightmost column shows the joint posterior distribution with 68\% (solid) and 95\% (dashed) contours. The true values are indicated by the star.}
\label{fig:llhdist}
\end{figure}

It is also possible to use $\mcl$ in a Bayesian approach. Using Bayes' theorem, the posterior
\begin{equation}
\prob(\vectheta|k) \propto \like(\vectheta|k) \pi(\vectheta),
\end{equation}
where $\pi(\vectheta)$ is a prior on the parameters. As evaluation of the normalization factor can by challenging, $\prob(\vectheta|k)$ can be approximated using a Markov Chain Monte Carlo (MCMC). For our toy example, we used \emcee{}~\cite{ForemanMackey:2012ig} to sample $\prob(\vectheta|k)$ under a uniform box prior for two different likelihood functions: $\mcl$ and $\adhoc$. The sampling was performed using the data and MC sets described in Sec.~\ref{sec:pointestimation}.

Figure ~\ref{fig:llhdist} shows the posterior distributions of $\Omega$ and $\Phi$. For each comparison, $\mcl$ (blue) and $\adhoc$ (orange) were sampled using the same underlying data and MC. We used 20 walkers with 300 burn-in steps followed by 1000 steps as settings for \emcee. The left and center column show the marginal posterior distribution for the mass, $\Omega$, and normalization, $\Phi$, respectively. The true value is indicated by the dashed, vertical line. The rightmost column shows the joint posterior distribution with 68\% (solid) and 95\% (dashed) contours. The true values are indicated by the star. With $\adhoc$, the true value of the parameter is highly improbable for the lower MC-size cases of the top and middle rows. In contrast, the posterior evaluated using $\mcl$ has increased width due to the reduced MC size. Even for $N_\mathrm{MC}=10^6$ (bottom row), the shape of the posterior evaluated using $\adhoc$ is narrower than that using $\mcl$. Credible regions estimated using $\adhoc$ would bias the result.

\subsection{Performance\label{sec:performance}}

\begin{figure}[ht]
\centering
    \includegraphics[width=1\linewidth]{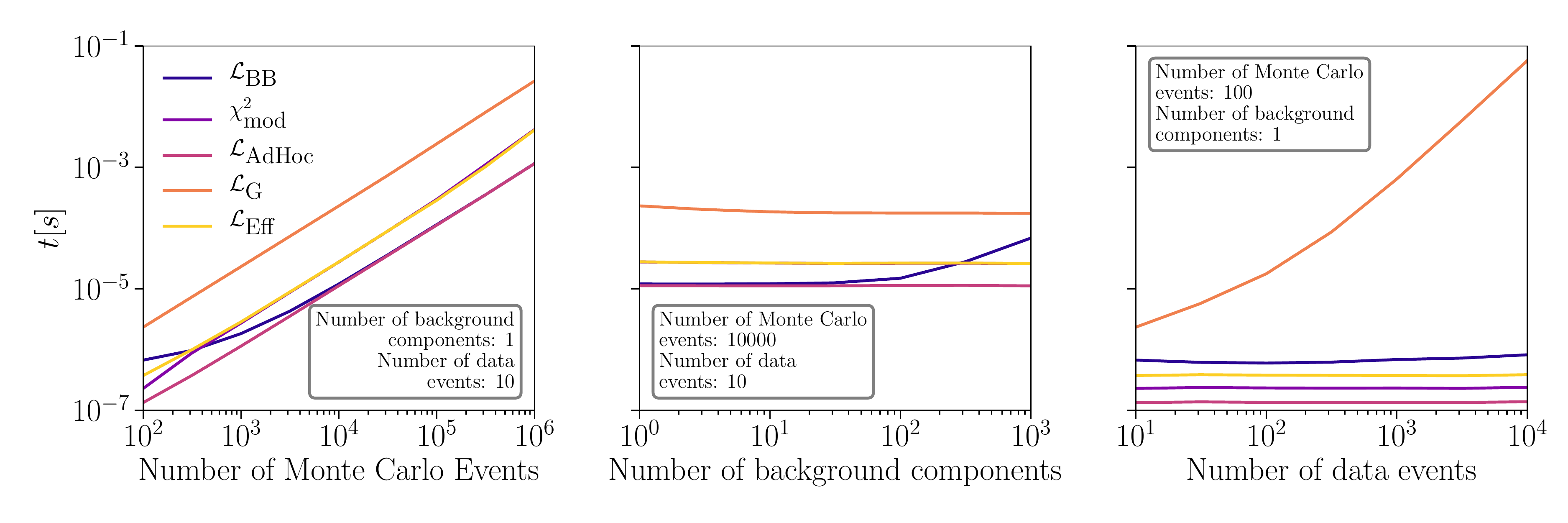}
\caption{\textbf{\textit{Likelihood function performance.}} Average single likelihood evaluation time is shown in the vertical axis in seconds. Different line colors show different likelihoods. Leftmost panel: the number of MC events used is shown on the horizontal axis. Center panel: the number of background components is shown on the horizontal axis. Rightmost panel: the number of data events is show on the horizontal axis.}
\label{fig:performance}
\end{figure}

In this section we compare our performance with other treatments available in the literature in terms of the runtime cost per likelihood evaluation for a single bin. We perform our tests using a single \texttt{Intel\textsuperscript{\textregistered} Core\texttrademark{} i5-8350U CPU @ 1.70GHz} running code compiled with \texttt{clang version 6.0.0-1ubuntu2}. We compute the likelihood CPU-evaluation time for the following likelihoods: $\adhoc$, modified-$\chi^2$, $\gl$~\cite{Glusenkamp:2017rlp}, $\lbarlow$~\cite{Barlow:1993dm}, and $\mcl$. For each of them we consider increasing number of MC events from $10^2$ to $10^6$, increasing number of background components from $1$ to $10^3$, and increasing counts of data events from $10^1$ to $10^4$. Figure~\ref{fig:performance} shows the behavior of the runtime with respect to these quantities. All likelihoods have runtime that increases with the number of MC events, as seen in the leftmost panel of Fig.~\ref{fig:performance}, as each likelihood must compute the sum of event weights which incurs an $\mathcal{O}(m)$ cost, where $m$ is the number of MC events in the bin. Additionally at low MC sample sizes the modified-$\chi^2$ is faster than $\mcl$ since $\mcl$ requires the evaluation of more expensive special functions, however at larger MC sample sizes this additional cost is negligible compared to that of summing the MC weights. In the middle panel of Fig.~\ref{fig:performance} it can be seen that all likelihoods except $\lbarlow$ are constant with respect to the number of background components as they only depend on summary statistics of the weight distribution. The Barlow-Beeston likelihood, $\lbarlow$, incurs an $\mathcal{O}(b d \log d)$ cost for solving a single root finding problem per physical component, where $b$ is the number of background components and $d$ is the number of digits of precision, and therefore is not constant in runtime with respect to the number of components. However, one key difference between $\lbarlow$ and $\mcl$ is that $\mcl$ must compute two summations (the sum of the weights and sum of the square weights), while $\lbarlow$ needs only to compute a single summation of the MC weights. The rightmost panel of Fig.~\ref{fig:performance} shows the runtime as a function of the number of data events; for most likelihoods the number of data events, $k$, enters only in the evaluation of some special functions which for all practical applications are approximately constant in runtime. $\gl$ evaluates a special function which for these purposes can only be computed in $\mathcal{O}(k^2 m)$ time, resulting in the dependence on the number of data events.
The $\adhoc$ treatment is always the fastest, but it does not incorporate MC statistical uncertainties in any way.  

\section{Conclusion\label{sec:conclusion}}

The use of MC to estimate expected outcomes of physical processes is nowadays standard practice. By construction, MC distributions are sample observations and subject to statistical fluctuations. MC events are also typically weighted to a particular physics model, and these weights may not be uniform across all events in an observable bin. 
A direct comparison of MC distributions to data is typically performed using $\adhoc$ or $\chi^2$, where the expectation from MC is computed as a sum over weights in a particular observable bin. Such likelihoods neglect the intrinsic MC fluctuations and may lead to vastly underestimated parameter uncertainties in the case of low MC size. A better approach is to use a likelihood that accounts for MC statistical uncertainties.

Along with the definitions of $\mu$ and $\sigma^2$ in Eq.~\eqref{eq:musigma}, the main result of this work is given in Eq.~\eqref{eq:parametrizedpoisson}. This new $\mcl$ is motivated by treating the MC realization as an observation of a Poisson random variate, computing the likelihood of the expectation using the MC and marginalizing the Poisson probability of observed data over all possible expectations. It is an analytic extension of the Poisson likelihood that accounts for MC statistical uncertainty under a uniform prior, $\prob(\lambda)$. By assuming that the number of MC events per bin is the outcome of sampling a Poisson-distributed random variable, and that the SPD is a good approximation of the CPD for arbitrary weights, $\like \left(\lambda|\vecw(\vectheta)\right)$ can be written in terms of $\mu$ and $\sigma^2$ as shown in Eq.~\eqref{eq:poisson_conditional_arb}. This allows us to calculate $\mcl$, given in Eq.~\eqref{eq:parametrizedpoisson}, and can be directly substituted in favor of $\adhoc$. Our construction is computationally efficient, exhibits proper limiting behavior, and has excellent coverage properties. In our tests, it outperforms other treatments of MC statistical uncertainty.

\section*{Acknowledgements} \label{sec:ack}

We thank Thorsten Gl\"usenkamp for useful discussions and Jean DeMerit for proofreading an early draft. CAA is supported by U.S. National Science Foundation (NSF) grant PHY-1505858. AS and TY are supported in part by NSF grant PHY-1607644 and by the University of Wisconsin Research Committee with funds granted by the Wisconsin Alumni Research Foundation.

\bibliographystyle{JHEP}
\bibliography{likelihood}

\newpage

\appendix

\ifx \standalonesupplemental\undefined
\setcounter{page}{1}
\setcounter{figure}{0}
\setcounter{table}{0}
\setcounter{equation}{0}
\fi

\renewcommand{\thepage}{Supplementary Methods and Tables -- S\arabic{page}}
\renewcommand{\figurename}{SUPPL. FIG.}
\renewcommand{\tablename}{SUPPL. TABLE}
\renewcommand{\arraystretch}{2}

\renewcommand{\theequation}{A\arabic{equation}}

\section{Summary of likelihood formulas\label{sec:appendixA}}

\begin{table}[h!]
\centering
\begin{tabular}{l | c}
\toprule
Parameters & $\mu \equiv \sum_{i=1}^m w_i, ~ \sigma^2 \equiv \sum_{i=1}^m w_i^2$ \\
\hline\hline
$\adhoc$ & $\frac{\mu^{k}e^{-\mu}}{k!}$\\ \hline
$\chi^2_{\rm mod}$ & $\frac{(k - \mu)^2}{\mu + \sigma^2}$ \\ \hline
$\lbarlow^{s=1}$ & $\underset{\bar m}{\rm max} \left\{
\frac{1}{k!m!} \left(\frac{\mu \bar m}{m}\right)^{k}\bar m^{m} e^{-\frac{\mu \bar m}{m} -\bar m} \right\}$ \\ \hline
$\meanl$ & $\left(\frac{\mu}{\sigma^2}\right)^{\frac{\mu^2}{\sigma^2}}\Gamma\left(k+\frac{\mu^2}{\sigma^2}\right)\left[k!\left(1+\frac{\mu}{\sigma^2}\right)^{k+\frac{\mu^2}{\sigma^2}}\Gamma\left(\frac{\mu^2}{\sigma^2}\right)\right]^{-1}$ \\ \hline
$\mcl$ & $\left(\frac{\mu}{\sigma^2}\right)^{\frac{\mu^2}{\sigma^2}+1}\Gamma\left(k+\frac{\mu^2}{\sigma^2}+1\right)\left[k!\left(1+\frac{\mu}{\sigma^2}\right)^{k+\frac{\mu^2}{\sigma^2}+1}\Gamma\left(\frac{\mu^2}{\sigma^2}+1\right)\right]^{-1}$ \\
\bottomrule
\end{tabular}
\caption{\textbf{\textit{Table of likelihood formulas.}} The likelihood functions discussed in this paper are given in each row. They are written in terms of $\mu$ and $\sigma$, whose explicit formulas are given in the top row, and the number of observed events, $k$, in the bin. In the case of $\lbarlow$ we write the likelihood for the single-process case. Our main result and recommended likelihood, $\mcl$, is given in the last row.}
\label{tbl:likelihoods}
\end{table}

\end{document}